\documentclass[10pt, conference, compsocconf]{IEEEtran}
\IEEEoverridecommandlockouts
\pdfoutput=1
\usepackage{fancyvrb}
\usepackage{cite}
\usepackage{amsmath,amssymb,amsfonts}
\usepackage{algorithmic}
\usepackage{graphicx}
\usepackage{textcomp}
\usepackage{xcolor}
\usepackage{listings}

\usepackage{comment}
\usepackage{pslatex}
\usepackage{tikz}
\usepackage{hyperref}
\usepackage{balance}
\usepackage{float}
\usepackage{stmaryrd}
\usepackage{multirow}
\usepackage{xspace}

\newcommand{\ifcond}{{\texttt{if}}\xspace}
\newcommand{\assumecond}{\textit{assume}\xspace}

\newcommand{\CCSP}{{$P^\prime$}\xspace}

\newcommand{\authnote}[2]{{\bf \textcolor{blue}{#1}: \em \textcolor{red}{#2}}}
\renewcommand{\authnote}[2]{}

\begin{document}
\title{
Benchmarking Symbolic Execution Using Constraint Problems - Initial Results
}

\author{\IEEEauthorblockN{Sahil Verma}
\IEEEauthorblockA{\textit{Department of Electrical Engineering} \\
\textit{IIT Kanpur, India}\\
%India \\
v.sahil1@gmail.com}
\and
\IEEEauthorblockN{Roland H.C. Yap}
\IEEEauthorblockA{\textit{School of Computing} \\
\textit{National University of Singapore, Singapore}\\
%Singapore \\
 ryap@comp.nus.edu.sg}
}

\maketitle

\begin{abstract}
Symbolic execution is a powerful technique for
bug finding and program testing.
It is successful in finding bugs in real-world code.
The core reasoning techniques use constraint solving, path exploration, and
search, which are also the same techniques used in solving combinatorial problems, e.g., finite-domain {\em constraint satisfaction problems} (CSPs). 
We propose CSP instances as more {\it challenging benchmarks}
to evaluate the effectiveness of the core techniques in symbolic execution. 
We transform CSP benchmarks into C programs suitable for testing the reasoning
capabilities of symbolic execution tools. 
From a single CSP $P$, we transform $P$ depending on transformation choice into different C programs.
Preliminary testing with the KLEE, Tracer-X, and LLBMC tools show substantial
runtime differences from transformation and solver choice.  
Our C benchmarks are effective in showing the limitations of existing symbolic execution tools.  
The motivation for this work is we believe that benchmarks of this form can spur the development and engineering of improved core reasoning in symbolic execution engines. 
\end{abstract}

%\begin{IEEEkeywords}
%symbolic execution, bounded model checking, constraint satisfaction problem, constraint solver
%\end{IEEEkeywords}

\section{Introduction}

Symbolic execution~\cite{king76symbolic} is one of the most powerful techniques for bug finding and program testing. It is successfully used on real-world code ~\cite{cadar08klee,godefroid05dart,sen05cute}. 
It is also used in reverse engineering \cite{chipounov10} and 
program repair \cite{nguyen13}.
Many real vulnerabilities have been found using either coverage-based
fuzzing, e.g., AFL \cite{AFL}, or symbolic execution.
In the Darpa Cyber Grand Challenge,
combining both fuzzing and symbolic execution was found to be effective
\cite{SOK-art-of-war,driller}.

Our goal is to help improve symbolic execution 
through benchmarking, in particular,
benchmarks which exercise the core reasoning techniques used 
which are essentially constraint solving, path exploration, and search.
% A known challenge for symbolic execution are programs when path exploration and constraint solving become more difficult
While path explosion \cite{cadar:sen:13,SOK-art-of-war} is a challenge for symbolic execution, we show that systematic variation of problem difficulty with benchmarks combining constraint solving and path exploration is a way of measuring the performance of a symbolic execution engine.
Constraint solving is also reported to be the dominant cost 
in symbolic execution of non-trivial programs 
\cite{trabish18chopped,liu14,palikareva13}.
Programs which naturally exhibit such challenges are when there is
difficulty to invert computation such as a cryptographic computation
or deliberately obfuscated code (including malware)
which has been written intentionally to be difficult
to analyze \cite{yadegari15}.
One intrinsic reason for this difficulty is that such code
is combinatorially difficult for symbolic execution (both path exploration
and/or constraint reasoning).
% Furthermore, when analyzing malicious code, then the malware authors
% can try to make analysis of the code more difficult for the tool by writing
% the code in a way which focuses on the kinds of code which the tool
% handles poorly.
% For example, by making the path exploration and constraint solving more
% combinatorially difficult, malicious behavior or intentional vulnerabilities
% are made more difficult to detect or analyze.

In this paper, we propose that the intrinsic path conditions challenges
which underlie symbolic execution can be
exercised and evaluated directly by using benchmarks of non-trivial 
combinatorial problems. 
% it is possible to evaluate how powerful is the core reasoning in a symbolic execution tool.
We are also motivated by how ``hard benchmarks'' from the SAT competition
have been used to drive the development of SAT solvers.
% It may be fair to say that the progress in SAT solvers may not have happened
% without the push from the SAT competition and SAT benchmarks.

There are also other practical issues which can challenge
symbolic execution such as system calls, memory modeling, heap reasoning \cite{cadar:sen:13,duck18}---these are orthogonal to this paper.
Our focus is on evaluating the core reasoning power of tools based
on symbolic execution independently from the other challenges above.
Unlike typical benchmarking of symbolic execution tools which use existing
programs, we design special {\it synthetic benchmarks} to deliberately ``stress test''
path conditions (constraint solving), path explosion, and exploration.

Symbolic execution relies on techniques for constraint
solving, typically \emph{satisfiability modulo theory} (SMT)
solvers, and handling of path conditions and path exploration.
Since the core techniques rely essentially on constraint solving and search, 
we propose the use of discrete combinatorial problems for
benchmarking. One well-known framework for discrete combinatorial problems
is \emph{Constraint Satisfaction Problems\/} (CSP).
We transform CSP instances to C programs to create our benchmarks. 
The transformation makes reaching a particular
program point the same as determining the satisfiability of the underlying CSP
(and also finding a solution). We give several transformations (into C
code) for the same CSP resulting in different C programs.
Ideally, as the CSP is the same, different tools should have similar runtimes.
The results show this is not the case,
justifying how our benchmarks can show weaknesses in the core
reasoning of the analysis tools.  
The benchmarks also test the sensitivity of symbolic execution to
different programs for the same problem
and scalability (where reasoning becomes more difficult).
% Our C benchmarks are open sourced~\cite{verma18cspc}.

Our preliminary experiments are on the following tools, which analyze
C programs using LLVM \cite{llvm17} as the common representation of
the program: KLEE \cite{cadar08klee} varying the constraint solvers
(STP \cite{ganesh07bit} and Z3 \cite{moura08z3}); Tracer-X \cite{tracerx17}
which adds interpolation to KLEE; and the LLBMC
\cite{falke13llbmc} bounded model checker
(LLBMC can be considered as ``static'' symbolic execution \cite{veritesting}).
% (Although LLBMC is not symbolic execution, it shares commonalities, see Sec \ref{sec:setup}).
The experiments show that
KLEE-based tools are much more sensitive than LLBMC to
the C constructs and structure of the code.  We also show that
the transformation (encoding) can affect different solvers in rather different ways.
% The benchmarks show that LLBMC generally executes the benchmarks faster. This result is understandable, given our benchmark focuses on the problems that are harder for a constraint solver, whereas the focus of the other symbolic execution tools is on the easier but many, which can be significant in practice. 
As our benchmarks involve intrinsically harder path exploration, 
preliminary experiments suggest that the execution times of the 
KLEE-based systems depend more
heavily on the solver used than on the sophistication of the search
the technique used, for instance, it seems that interpolants and abstraction
learning does not yet have sufficient gains.
% Overall, our benchmarks show that the reduction in the
% number of paths explored, which in turn reduces the number of solver
% calls does help performance. 
There are also substantial
differences in running times by merely replacing the solver
used for the same tool.

The paper is organized as follows.
Related work is presented in Section \ref{sec:related} and 
background on CSP in Section
\ref{sec:background}.
The transformation of CSP benchmarks into C programs is given
in Section \ref{sec:transformation}. Experimental results on the C
benchmark suite are given in Section \ref{sec:results} and  Section \ref{sec:conclusion} concludes.

\section{Related Work}
\label{sec:related}

Symbolic execution generalizes program instructions into
mathematical relations (\emph{constraints\/}) between the state before
and after each instruction~\cite{king76symbolic}. In this way, instead
of actually executing each instruction, a symbolic execution system
collects the constraints that are translated from each instruction
into a conjunction, called the \emph{path condition}, which can be
tested for satisfiability using constraint solvers, e.g.
\emph{satisfiability modulo theories\/} (\emph{SMT\/})
solvers~\cite{moura11}, such as STP~\cite{ganesh07bit} or
Z3~\cite{moura08z3}. Symbolic execution is widely used for
\emph{whitebox\/} fuzzing, the advantage is that each program path is
only tested once, whereas traditional black-box fuzzing may exercise
the same path multiple times.  

KLEE~\cite{cadar08klee, KLEEgit, KLEEpapers} is one of the most
well-known symbolic execution tools and shown to be effective.
% \footnote{ As we evaluate KLEE-based tools, readers unfamiliar with symbolic execution may find the KLEE tutorial \cite{KLEEtutorial} to be helpful. }
A critical challenge in symbolic execution is to make it more scalable.
One approach is Tracer-X~\cite{tracerx17} which extends
KLEE with learning (interpolation) 
to prevent redundant execution~\cite{jaffar09intp} to mitigate
path explosion.
A recent work, {\em chopped symbolic execution}~\cite{trabish18chopped}, 
also reduces path explosion by extending KLEE so that
parts of the code can be manually excluded by executing them lazily during
symbolic execution to make overall execution faster.  
In our benchmarks,
the programs are transformations of non-trivial CSP benchmarks such that
there is no obvious part of the code, i.e., the transformed
C program, that can be removed from consideration.  
Thus, chopped symbolic execution is not applicable.
In this paper, our benchmarks are C code, hence, we focus on KLEE-based systems. 

Other approaches related to symbolic execution are
DART~\cite{godefroid05dart}, CUTE~\cite{sen05cute},
SAGE~\cite{godefroid12sage}, and S$^2$E~\cite{chipounov11s2e}.
These can be
categorized as \emph{concolic\/} testers, where the construction of
the path condition follows an actual execution of the program. 
Surveys of symbolic execution are in
\cite{baldoni18symbolic,cadar:sen:13}.
% A survey of symbolic execution is in Baldoni et al. \cite{baldoni18symbolic} and \cite{cadar:sen:13}

There has been work to measure the capabilities of symbolic execution
tools.  Several challenges are identified with concolic testing, in
particular, constraint solving~\cite{kannavara15concolic}.
Benchmarking of symbolic execution tools with \emph{logic bombs}, 
code fragments which can cause difficulty for symbolic execution such
as buffer overflow, arithmetic overflow, etc.
% code fragments that can only be executed when certain conditions are met, 
is proposed in \cite{xu18logicbombs}.  
We differ in that we address the challenge which occurs in the simplest
code, this is different from challenges which can arise from dealing
with undefined behavior in C such as buffer overflow.
Another challenging problem 
for symbolic execution is how to handle
cryptographic functions \cite{xu17cryptographic,corin11efficient},
the combinatorial challenges are related to the CSP benchmarks we propose.
Evaluation of symbolic execution tools applied to industrial
software is in \cite{qu11case} and considers limitations in
handling language features. 
Language features are also tested in \cite{cseppento15evaluating} 
using Java and .NET test generators.

\section{CSP Background}
\label{sec:background}

One formulation of constraint problems are Constraint
Satisfaction Problems (CSP).
A {\em Constraint Satisfaction Problem} (\emph{CSP\/}) $P$ is a pair $(X , C)$ where $X$ is a set
of $n$ variables $\{x_1, . . . , x_n\}$ and $C$ a set of $e$ constraints
$\{c_1, . . . , c_e\}$ \cite{dechter03constraint,cp-handbook}. 
In this paper, we focus on finite discrete combinatorial problems.
In a {\em finite domain CSP}, 
variables $x \in X$ take values from their domain $D(x)$ which is
a finite set of values.
% We use $(x, a)$ to denote the value $a\in D(x)$ (or simply $a$ when the context is clear). 
Each $c\in C$ has two components: a
scope ($scp(c)$) which is an ordered subset of variables of
$X$; and a relation over the scope ($rel(c)$). Given $scp(c) =
{x_{i1}, . . . , x_{ir}}, rel(c)\subseteq \Pi^r_{j=1}D(x_{ij})$ is 
the set of satisfying {\em tuples} (combinations of values)
 for the variables in $scp(c)$.
% We also refer to $c$ by $c(x_{i1}, . . . , x_{ir})$ and a {\em tuple} of $c$ is $(a_{i1}, . . . , a_{ir}) \in rel(c)$.
% The {\em arity} of a constraint $c$ is the number of variables in $c$, $|scp(c)|$.
A constraint is {\em satisfied} if there is valuation for the variables 
from at least one of the tuples in
its relation which takes values from the variables' domain.
A {\em solution} to $P$ is a valuation for $X$ such that every constraint
is satisfied. $P$ is {\em satisfiable} iff a solution exists.
If no solution exists then $P$ is {\em unsatisfiable}.

% Note that the general form of a finite domain CSP is NP-complete.
% More CSP background can be found in \cite{dechter03constraint,cp-handbook}.

\begin{figure}[t]
  {\small
\begin{verbatim}
<group>
<extension>
<list> %0 %1 %2 </list>
<conflicts> (0,0,0) (0,1,0)</conflicts>
</extension>
<args> x[0] x[1] x[2] </args>
<args> x[3] x[4] x[5] </args>
</group>
\end{verbatim}}
\vspace{-4mm}

\center{\small(a) Extensional constraint example with 0-1 Domains}
\vspace{-1mm}

{\small
    \begin{verbatim}
<group>
<intension> eq(%0,dist(%1,%2)) </intension>
<args> y[0] x[0] x[1] </args>
<args> y[1] x[1] x[2] </args>
</group>
\end{verbatim}}
\vspace{-6mm}

\center{\small(b) Intensional constraint example ($x = |y - z|$)}
\vspace{-1mm}
    
{\small
  \begin{verbatim}
    <allDifferent> x[0] x[1] x[2] </allDifferent>
\end{verbatim}}
\vspace{-4mm}

\center{\small(c) Alldifferent (intensional) constraint}
\vspace{-2mm}

\caption{Example CSP Constraints in XCSP3 Format}
\label{fig:csp-example}
\vspace{-16pt}
\end{figure}

\begin{table*}[tb]
  \begin{center}
\begin{tabular}{|c|c|c|c|c||c|c|c|c|} \hline
	\textbf{Type} & \textbf{Version} & \textbf{Construct} & \textbf{Operator} & \textbf{Grouped} & \multicolumn{3}{|c|}{\bf Benchmarks (\#{\it lines}, \#{\it variables})} \\
	  & & & & &
	\textbf{Sat-Aim100} & \textbf{Sat-Aim200} & \textbf{Dubois}\\ \hline\hline
\multirow{12}{*}{\rotatebox[origin=c]{270}{Extensional}}
& 1      & \ifcond    & \textit{logical}   & \textit{no}  & (853, 100) & (2080, 200) & (933, 130) \\
& 2      & \ifcond    & \textit{logical}   & \textit{yes} & (352, 100) & (657, 200) & (432, 130)\\
& 3       & \ifcond    & \textit{logical}   & \textit{all} & (316, 100) & (616, 200) & (412, 130) \\
& 4         & \ifcond    & \textit{bitwise}   & \textit{no} & (1040, 100) & (2080, 200) & (933, 130)\\
& 5       & \ifcond    & \textit{bitwise} & \textit{yes} & (352, 100) & (657, 200) & (432, 130) \\
& 6        & \ifcond    & \textit{bitwise} & \textit{all} & (316, 100) & (616, 200) & (412, 130) \\
& 7       & \assumecond & \textit{logical}   & \textit{no} & (782, 100) & (1552, 200) & (759, 130) \\
& 8     & \assumecond & \textit{logical}   & \textit{yes} & (531, 100) & (1033, 200) & (676, 130) \\
& 9      & \assumecond & \textit{logical}   & \textit{all} & (516, 100) & (1016, 200) & (672, 130) \\
& 10       & \assumecond & \textit{bitwise} & \textit{no} & (782, 100) & (1552, 200) & (759, 130) \\
& 11     & \assumecond & \textit{bitwise} & \textit{yes} & (531, 100) & (1033, 200) & (676, 130) \\
& 12     & \assumecond & \textit{bitwise} & \textit{all} & (782, 100) & (1552, 200) & (759, 130) \\ \hline
\multirow{10}{*}{\rotatebox[origin=c]{270}{Intensional}}
& &&&& \textbf{AllInterval} & \textbf{CostasArray} & \textbf{HayStacks}\\ \hline
& 1    & \ifcond    & \textit{NOP}   & \textit{no}  & (549, 74) & (718, 98) & (3302, 173) \\
& 2       & \ifcond    & \textit{logical}   & \textit{yes} &  (259, 74) & (361, 98) & (555, 173) \\
& 3        & \ifcond    & \textit{logical}   & \textit{all} &  (239, 74) & (310, 98) & (535, 173) \\
& 4        & \ifcond    & \textit{bitwise} & \textit{yes} &  (252, 74) & (354, 98) & (548, 173) \\
& 5         & \ifcond    & \textit{bitwise} & \textit{all} &  (238, 74) & (309, 98) & (534, 173) \\
& 6     & \assumecond & \textit{NOP}   & \textit{no} &  (2727, 74) & (1189, 98) & (2231, 173) \\
& 7      & \assumecond & \textit{logical}   & \textit{yes} &  (242, 74) & (323, 98) & (538, 173) \\
& 8      & \assumecond & \textit{logical}   & \textit{all} &  (238, 74) & (309, 98) & (534, 173) \\
& 9      & \assumecond & \textit{bitwise} & \textit{yes} &  (242, 74) & (323, 98) & (538, 173) \\
& 10      & \assumecond & \textit{bitwise} & \textit{all} &  (238, 74) & (309, 98) & (534, 173) \\ \hline
\end{tabular}
\end{center}
\caption{Transformation Versions and Features Used. 
Details of transformations and features are on the left of the ($||$).
Benchmark statistics are on the right of ($||$): 
average number lines of the C code and 
number of variables of the CSP instances.
}
\label{table:features}
\end{table*}

A constraint can be defined in two ways:
(i) intensional; or (ii) extensional (also known as {\em table constraint}).
An {\em intensional constraint} is one where the relation of the
constraint is defined implicitly. This is a common form
of constraints supported in constraint solvers, e.g., a linear arithmetic inequality
is implicitly understood in the usual arithmetic fashion over either
real numbers or integers.
Constraints can also be defined {\em extensionally}---the relation is
defined as a (positive) {\em table} giving the satisfying tuples of 
constraint $c$ (respectively, also a
``{\em negative table}'' defining the set of ``{\em negative tuples}'', namely,
the tuples not satisfying the constraint).
Consider the constraint $x + y = 1$ where the domain of the variables
is binary.
The equation above is already in the intensional form of the constraint. 
The positive extensional form is given by the tuples 
$\{\langle 1,0\rangle, \langle 0,1\rangle\}$ 
(respectively, the negative extensional form is
$\{\langle 0,0\rangle, \langle 1,1\rangle\}$)
for the 
variable sequence $\langle x,y \rangle$.

As our benchmarks used the XCSP3 format, we briefly describe XCSP3.
Figure \ref{fig:csp-example} illustrates three kinds of constraints in
XCSP3 format~\cite{boussemart16xcsp3,xcsp18}.
XCSP3 is an XML format for defining CSP instances which can be used
with many constraint solvers.
It is also used in the XCSP3 competition for constraint solvers
\cite{xcsp:competition}.
We use XCSP3 problem instances in our experiments.
The details of the XCSP3 format are beyond of the scope
of the paper, so examples in XCSP3 are meant to be illustrative only.

We consider CSPs where the constraints are in: 
(i) intensional form, and
(ii) extensional form 
(see Figure \ref{fig:csp-example} (a) which defines a negative table 
with two constraints
on $\langle x0, x1, x2 \rangle$ and $\langle x3, x4, x5 \rangle$ using
the same table definition).
Some examples of intensional constraints are:
\verb|eq(%0,dist(%1,%2))|
representing $x = |y - z|$ for a given substitution of $x$, $y$, and
$z$ (see Figure \ref{fig:csp-example} (b)); and the \textit{alldifferent\/}
constraint (see Figure \ref{fig:csp-example} (c)), which constrains the variables 
$\langle x0, x1, x2 \rangle$
to all take different values.
We remark that {\it alldifferent} is considered a {\em global constraint} but
in this paper, what matters is the intensional versus extensional distinction
of the constraint.

\definecolor{myc}{RGB}{0, 150, 0}
\lstset{
  basicstyle=\small\ttfamily,
  xleftmargin=4mm,
  mathescape=true,
  escapeinside=`'
}

\begin{figure}[tb]
%\begin{Verbatim}[fontsize=\small,xleftmargin=5mm]
\begin{lstlisting}
`\textcolor{myc}{int x0, x1, x2, x3, x4, x5;}'
//  `\it declare variables symbolic'
`\textcolor{myc}{klee\_make\_symbolic(\&x0,sizeof(x0),"x0");}'
`\textcolor{myc}{klee\_make\_symbolic(\&x1,sizeof(x1),"x1");}'
  $\cdots$
`\textcolor{myc}{klee\_make\_symbolic(\&x5,sizeof(x5),"x6");}'
//  `\it enforce variable domains'
`\textcolor{myc}{klee\_assume(x0 $\geq$ 0 \&\& x0 $\leq$ 1);}'
`\textcolor{myc}{klee\_assume(x1 $\geq$ 0 \&\& x1 $\leq$ 1);}'
  $\cdots$
`\textcolor{myc}{klee\_assume(x5 $\geq$ 0 \&\& x5 $\leq$ 1);}'
if ((x0==0 && x1==0 && x2==0) ||
    (x0==0 && x1==1 && x2==0)) exit(0);
if ((x3==0 && x4==0 && x5==0) ||
    (x3==0 && x4==1 && x5==0)) exit(0);
`\textcolor{myc}{assert(0);}' // `\it CSP is satisfiable'
\end{lstlisting}
%\end{Verbatim}

\vspace*{-2mm}
\center{\small(a) Extensional Version 1}

\vspace*{-1mm}
%\begin{Verbatim}[fontsize=\small,xleftmargin=5mm]
\begin{lstlisting}
if ((x0==0 & x1==0 & x2==0) | 
    (x0==0 & x1==1 & x2==0) |
    (x3==0 & x4==0 & x5==0) |
    (x3==0 & x4==1 & x5==0)) exit(0);
\end{lstlisting}
%\end{Verbatim}
\vspace*{-4mm}
\center{\small(b) Extensional Version 5}

\vspace*{-1mm}
%\begin{Verbatim}[fontsize=\small,xleftmargin=5mm]
\begin{lstlisting}
klee_assume(!((x0==0 && x1==0 && x2==0) ||
              (x0==0 && x1==1 && x2==0) ||
              (x3==0 && x4==0 && x5==0) ||
              (x3==0 && x4==1 && x5==0)));
\end{lstlisting}
%\end{Verbatim}
\vspace*{-4mm}
\center{\small(c) Extensional Version 8}

\vspace*{-1mm}
%\begin{Verbatim}[fontsize=\small,xleftmargin=5mm]
\begin{lstlisting}
if (y0==dist(x0,x1)); else exit(0);
if (y1==dist(x1,x2)); else exit(0);
\end{lstlisting}
%\end{Verbatim}

\vspace*{-4mm}
\center{\small(d) Intensional Version 1 (dist)}

\vspace*{-1mm}
%\begin{Verbatim}[fontsize=\small,xleftmargin=5mm]
\begin{lstlisting}
if ((y0==dist(x0,x1) && y1==dist(x1,x2)));
  else exit(0);
\end{lstlisting}
%\end{Verbatim}

\vspace*{-4mm}
\center{\small(e) Intensional Version 2 (dist)}

\vspace*{-1mm}
%\begin{Verbatim}[fontsize=\small,xleftmargin=5mm]
\begin{lstlisting}
klee_assume(x0!=x1 & x0!=x2 & x0!=x3)
\end{lstlisting}
%\end{Verbatim}
\vspace*{-4mm}
\center{\small(f) Intensional Version 9 (alldifferent)}

\vspace*{-1mm}
%\begin{Verbatim}[fontsize=\small,xleftmargin=5mm]
\begin{lstlisting}
if (x0!=x1 && x0!=x2 && x0!=x3 &&
   y0==dist(x0,x1) && y1==dist(x1,x2)) 
   assert(0);
\end{lstlisting}
%\end{Verbatim}
\vspace*{-4mm}

\center{\small(g) Intensional Version 3 (dist and alldifferent)}
\caption{Example transformed constraints with version numbers from Table \ref{table:features}. Version 1 is the full transformation, the others are
	only the constraint.
}
\label{fig:transformation}
\vspace*{-8mm}
\end{figure}

\section{Transforming Combinatorial Problems to C}
\label{sec:transformation}

The idea behind turning a CSP into a C program is as follows.
Solving a combinatorial problem can be imagined as two components:
firstly an oracle guesser for a solution; and
secondly a checker for the solution.  
We can think of symbolic execution as finding a solution 
which the oracle could have returned 
and we transform the CSP into a C program that functions as the checker.

Our approach is to transform a combinatorial problem, a CSP $P$,
into a C program \CCSP to be tested on program analysis tools as follows:
(i) the finite domain variables of the CSP correspond to integer variables 
in the program \CCSP (C variables are treated as symbolic);
(ii) CSP variable domains are converted to \emph{assume} statements (see below);
and
(iii) the constraint relations are encoded into conditional statements in
the program or as \emph{assume\/} statements.
The encoding of the CSP to C ensures that when the CSP $P$
is satisfiable (symbolic) execution of \CCSP is able
to reach a {\it distinguished program point}---the values of 
the C variables in \CCSP are a solution to CSP $P$.
% \footnote{ KLEE can construct a test case, which is another way of checking that the variable values in the test case satisfy the constraints in $P$, which can be useful for checking the encoding.  }
To test the analysis tools, the distinguished program
point is mapped to an assertion failure, i.e. \verb|assert(0)|.
Similarly, when CSP $P$ is unsatisfiable, the assert cannot be reached from
any path, i.e., all paths will not succeed.

From a single CSP $P$, we propose
several transformations of $P$ into C programs.% corresponding to $P$.
% a number of 
% 11 transformations for CSPs with extensional constraints and 12
% transformations for CSPs with intensional constraints.  
The purpose of various transformations which result in different programs is to exercise the tools in different ways (as we show in the results).  
We employ two general approaches for encoding the constraints in $P$:
% (the tools we evaluate are KLEE, Tracer-X and LLBMC, see also Sec \ref{sec:setup})
\begin{enumerate}
\item \textbf{\ifcond} approach: \texttt{if} statements are
  created, whose condition captures the values of variables satisfying the constraint. The execution can terminate with an
  \texttt{exit(0)} statement in their \emph{else\/} branches. If it is possible that execution takes the \emph{then\/} branches, then there will be satisfying values for the variables of the encoded constraint. In symbolic execution, the condition is simply
  added into the current set of constraints, i.e., the \emph{path
    condition}.
    % \RY{combine ifcond and ifneg}
    % Although we have explained that the valuation not satisfying the
    % condition is the {\tt then} branch, an alternative is to invert it with the
    % condition representing satisfaction, in which case the {\tt exit(0)} will
    % be in the {\tt else} branch.
    % \RY{can use this to have a {\it if*} case for case 1 intentional in Table 1}

%\item \textbf{\ifneg} approach. Similar to the \ifcond approach, but
%  inversely, the \emph{then\/} branch of the if conditional represents
%  the satisfaction of the constraint, whereas the \emph{else\/}
%  represents branch the opposite. Therefore, here we place
%  \texttt{exit(0)} statement at the \textit{else\/} branches of the
%  \texttt{if} statements instead of the \textit{then\/} branches. The
% constraint is thus added into the path condition only when the
%  \textit{then\/} branch is executed.

\item \textbf{\textit{assume\/}} approach: the constraint
  solver of the analysis tool is used directly---the constraints of
  the CSP $P$ are translated into an argument to \emph{assume\/}
  statements (\texttt{klee\_as\-sume} of KLEE or
  \texttt{\_\_llbmc\_assume} of LLBMC). When an assume statement is executed in symbolic execution, the constraint is simply added into the path condition and not intended to be executed as C, whereas the \textbf{\ifcond} is C code.
  %% At the end of the execution, we evaluate the
  %% path condition using the backend SMT solver.
\end{enumerate}

When all constraints have been translated, we place an \texttt{assert(0)}
at the end of the program (the distinguished program point). 
The \texttt{assert(0)} triggers an
assertion failure, so if the CSP $P$ is satisfiable (no contradiction
is found), the program terminates with an assertion failure. 
The generated test case from symbolic execution, e.g., KLEE or Tracer-X,
is a solution for $P$. 
If the CSP $P$ is unsatisfiable,  assertion failure will
not be possible and will be reported as a failure.

Table \ref{table:features} (to the left of the double vertical line) 
shows how different features are combined
to obtain different transformed (versions) C programs for a CSP $P$. 
Overall we have designed 12 extensional transformations for extensional CSPs 
and 10 intensional transformations for intensional CSPs.\footnote{
%Not all operators are needed for intensional constraints such as the 
%NOP operator.
% (see {\bf Operator} section).
Intensional versions do not need operators in every case (shown as NOP in the table).
For example, both Extensional Versions 1 and 4, correspond to Intensional Version 1.
%and both Extensional Versions 7 and 10 correspond to Intensional Version 6. 
% (12 Vs. 10).
This explains the difference in the number of transformations for extensional and intensional problems.
}
Each transformation version (different row in Table \ref{table:features}) employs different C constructs on the same underlying CSP $P$.  
% We explain the transformations using selected examples in Figure \ref{fig:transformation} which cover the main transformation choices.  
The transformations are designed to be correct by construction. 
There is insufficient space to formalize the transformations, rather,
we do it by example.
Figure \ref{fig:transformation} (a) shows how
the negative tables in Figure \ref{fig:csp-example} (a) 
are transformed into C under the Version 1 transformation. The \textcolor{myc}{green} coloured code in Figure \ref{fig:transformation}(a) shows details of the transformation common to all versions of a particular CSP: C variable declaration, KLEE symbolic variable construction, constraining variable domain, and distinguished assertion program point.
CSP variables are symbolic in KLEE and domains are encoded 
with \verb|klee_assume| (similarly for LLBMC).
Reaching the \verb|assert| means the CSP is satisfiable.
Similarly, Figures \ref{fig:transformation} (b),
and (c) are the results of Extensional Versions 5 and 8 on the same constraint.
% respectively on Figure \ref{fig:csp-example} (a).
Figure \ref{fig:transformation} (d) and (e) are from Intensional
Versions 1 and 2 on Figure \ref{fig:csp-example} (b), and Figure
\ref{fig:transformation} (f) is from Intensional Version 9
on Figure \ref{fig:csp-example} (c). Figure
\ref{fig:transformation} (g) is from Intensional Version 3
combining constraints in Figures \ref{fig:csp-example} (b) and (c) 
as a single problem.

Table \ref{table:features} shows how the transformations in 
each column (Construct, Operator, Grouped) are
combined to give a particular version.
% The transformations summarized in Table \ref{table:features} are as follows.  
The \textbf{Construct} column gives the
construct used in the C program, if/assume approaches.  Figures
\ref{fig:transformation} (a) and (b) are the transformation results
of the \ifcond approach on the constraint example in Figure
\ref{fig:csp-example} (a),
% \footnote{ As Figure \ref{fig:csp-example} (a) is given as a negative table with negative tuples, the C code for matching negative tuples in Figure \ref{fig:transformation} (a) exits. }
 while Figure
 \ref{fig:transformation} (c) shows the result of the
 \textit{assume\/} approach, which uses an assume statement
 (example is for KLEE using \texttt{klee\_assume}, whereas when
 testing on LLBMC, \texttt{\_\_llbmc\_assume} is used).
%  (\texttt{klee\_assume} for KLEE and \texttt{\_\_llbmc\_assume} for LLBMC).
 
% This can be added. 
%  and finally
%  Figure \ref{fig:transformation} (e)
%  shows the result of the \ifneg\/ approach on Figure \ref{fig:csp-example} (b).

%% \ifneg denotes the use of ``if conditionals''
%% with negated conditions, 
%% \RY{text needs to be revised}
%% essentially this gives a possible path for the constraint to fail with
%% values for the variables.
%% If the path continues, then the constraint has satisfiable values for it's
%% variables.
% which either execute \texttt{assert(0)} in the then branch or execute \texttt{exit(0)} in the else branch instead of the then branch. 
% \vspace{-3.5mm}

The \textbf{Operator} column gives the logic operators
used in the conditions, C logical operators (\verb|&&|
or \verb+||+) or C bitwise operators, which are used to combine the
expressions making up the conditions.
The difference between the 
logical and bitwise operators is that the logical
operators are from C-style \emph{short-circuiting} of conditions, i.e. 
the condition is broken up into a cascading conditional branching
structure with only a \emph{atomic\/} condition guarding every
branch. Atomic conditions do not contain logical or bitwise operators.
%% \footnote{ The logical
%%   operators have short-circuiting of the expression while bitwise
%%   operators do not.  }
Figure \ref{fig:transformation} (b) and (f) % \RY{added f}
shows the usage
of bitwise operators in the transformation, while 
Figure \ref{fig:transformation} (a), (c), (e) and (g) show the usage of logical operators.
% \RY{fixed version number}
% Here, N/A is given for the
% intensional constraints transformation Version 1 since the assume
For Intensional Transformations, when there is no grouping of conditions, 
no operator is needed (denoted by \emph{NOP} in Version 1 and 6) in Table \ref{table:features} and illustrated in Figure \ref{fig:transformation} (d).
% \RY{changed to 2d}
% statements
% only use atomic constraints, hence, no logical operator is used,
% which is denoted by N/A.  
The CSP may have several constraints defined in the same way but on
different variables, i.e. in Figure \ref{fig:csp-example} (a),
there is a single constraint group with two individual constraints.

% \vspace{-3.5mm}

The \textbf{Grouped} column shows 
how translation is grouped by constraints:
 the translation is per constraint group (\textit{yes}); per
 individual constraints defined in the group (\textit{no});
%  by $\TagTerminal{arg}$ (see examples in Figure \ref{fig:csp-example}) within the constraint group (\textit{no}), 
 or the entire CSP $P$ is grouped together as a single condition
 (\textit{all}).  
For example, Figure \ref{fig:transformation} (a) and (d)
 show the \textit{no\/} case, 
Figure \ref{fig:transformation} (g) shows the \textit{all\/} 
case given the
constraints in Figures \ref{fig:csp-example} (b) and (c),
 and the rest of Figure \ref{fig:transformation}
  show the \textit{yes\/} case.
 Figure \ref{fig:transformation} (g) simply combines
 the \texttt{assert(0)} statement in
 the \textit{then\/} body of the \texttt{if} statement.
%% The \textbf{Storage} column shows how variables are
%%  declared in the C program. For most transformations, we declare the
%%  variables to be stack variables,
%% % on the local stack of C's \texttt{main} function (denoted as \textit{stack\/}),
%% but for some we experimented
%%  with using heap allocation (denoted as \textit{heap\/}).
% Our benchmarks using the transformations are open sourced~\cite{verma18cspc}.
We remark that as the benchmarks are for testing path explosion and constraint
solving, the transformations produce straight-line code.
However, it is straightforward to make more compact forms with loops as well, 
but that only makes symbolic execution more complex.
\enlargethispage{0.5\baselineskip}

\section{Experiments}
\label{sec:results}

\subsection{Experimental Setup}
\label{sec:setup}

We present experiments evaluating selected symbolic execution tools
on the C programs created from
our transformations of extensional and intensional CSP problems
from the XCSP3 benchmarks~\cite{xcsp18}.
We tested three classes of extensional constraint problems:
% \RY{add Dubois}
AIM-100 and AIM-200
(each consists of 16 satisfiable and 8 unsatisfiable problems),
and Dubois (30 unsatisfiable problems).
% \RY{add stats?}).
%\RY{NOTE: remember to add text for Dubois results later}
%\SV{Only putting scalability plot for Dubois, not the robustness plot (space issues).}
For intensional CSP problems, we used
the All-Interval (19 satisfiable problems), Costas-Array (8
satisfiable problems), and Haystacks (14 problems)
% \footnote{ The satisfiability was not listed for Haystacks instances. } 
problem classes.  
The intensional constraints used in these CSP instances
are shown in Table \ref{table:constraints}.
Statistics on the C benchmarks are given on the right of ($||$)
in Figure \ref{fig:transformation} giving the average number of
lines of code in the C programs and the average number of variables 
of the CSP instance for the problem classes tested.
These problem instances from the XCSP3 benchmarks were chosen primarily because the transformation results are comparatively smaller
than other instances (e.g., most other extensional XCSP3 benchmarks gave much larger C programs which could not finish running in a day).
% \footnote{ For example, most other extensional XCSP3 benchmarks led to much larger C programs whose runtimes were more prohibitive for the symbolic execution tools. }
% The rest of problem instances are satisfiable. 
%The focus of our experiments is to identify the
%points for improvements in symbolic execution using our benchmarks.
%
All C programs in our benchmark suite are compiled into common
bytecode (LLVM bitcode) instances using LLVM's \texttt{clang} at \texttt{-O3}. 
Our benchmarks are publicly available~\cite{verma18cspc}.
% We used \texttt{-O3} flag because the optimized files gave us much better timing than unoptimized ones. 

Experiments were run on a 3.40 GHz Intel i7-6700 32GB RAM machine
running Ubuntu 16.04.
We focus on C code and only evaluate tools which are available for C.
To keep things on a common footing, we experimented with
the following LLVM-based tools analyzing the 
{\it common compiled C instances}: KLEE \cite{cadar08klee}
with STP and Z3 as the backend SMT solvers, Tracer-X  \cite{tracerx17}
which uses only Z3 as the backend solver, and LLBMC, which uses
STP as the backend solver.  
KLEE is used as it is perhaps the most well-known symbolic
execution engine
and is commonly used, at least 130 papers use KLEE \cite{KLEEpapers}.
It was 2nd placed, highest systematic tool, in the 2019 Test-Comp competition.
%\footnote{ For example, the recent Chopped symbolic execution ~\cite{trabish18chopped} only compares with KLEE. }
Tracer-X builds upon Tracer \cite{jaffar12},
extending KLEE with abstraction learning \cite{jaffar99} and
interpolants \cite{craig57} which can prune paths to mitigate 
the path explosion problem.
LLBMC \cite{falke13llbmc} is not a symbolic execution engine, it
is a bounded model checker, but it also addresses the same core issues
(constraint encoding, constraint solving, solver search space)
and uses the same underlying solvers (STP). 
The results show LLBMC to be a relevant and useful comparison.
All tools used a timeout of 1000 seconds.\footnote{
LLBMC exits upon finding a solution, for a fairer comparison, we
use the option \texttt{--exit-on-error} so that KLEE and Tracer-X also
exit upon reaching \texttt{assert(0)} statement (cf. Section
\ref{sec:transformation}) at the end of the program. 
}

%We note that
%being bounded model checking, LLBMC allows the specification of loop
%and recursion depths, but since our benchmarks are neither looping nor
%recursive, these are irrelevant.  

We also ran the AbsCon CSP
solver~\cite{merchez01abscon,xcsp-tools} with default options
on the original CSP instances as a baseline comparison for the XCSP3 instance.
% to serve as one baseline comparison with the tools.
AbsCon is a Java-based constraint solver using the
XCSP3 format. As such, no further input encoding or transformations are needed for XCSP3 instances,
making it a convenient comparison.
We use the AbsCon baseline to make it easy to present the timings---normalizing the execution time of the tools to the AbsCon timing.\footnote{
AbsCon is not meant for solver comparison; interestingly, some LLBMC timings are faster.
}
%% This serves as
%% one baseline to make it easier to compare the execution time of the
%% tools on the various transformed C programs by normalizing to the
%% AbsCon timing.
% AbsCon was run with the default options.
%% via the command \texttt{java}
%% \texttt{AbsCon.Resolution} \textit{filename\/}.
%\AS{AIM-100 LLBMC: 0.15 to 0.22, AIM-200: 0.17 to 0.26}

We evaluate the benchmarks on the analysis tools on:
(i) transformation \emph{robustness\/}; and (ii) tool \emph{scalability}. 
Transformation robustness measures how varying the transformations
used on a single problem instance $P$ changes the execution time.
Ideally, if the tool is robust with respect to the transformations, the
timing variations should be small, i.e., it is insensitive.
We highlight that the different transformed C programs are on the same
instance $P$, hence, the programs can be ``considered equivalent''.
One might expect (or even demand) a certain level of robustness for
equivalent code. However, the results show that the tools are not robust
with respect to transformation.

Tool scalability is the change in tool analysis time as the 
size of the problem changes 
(the chosen scalability benchmarks have a size parameter).
In the scalability experiments, 
varying the size naturally changes the problem instance but the 
problem and code structure remains similar.
We remark that although the AIM problems vary the number of variables,
we do not use them for scalability as
they do not have a structural scaling parameter defined
by the problem, i.e. the number of variables may not be a fair parameter.

\begin{table}[t]
  \center{\begin{tabular}{|c|c|} \hline
    \textbf{Problem} & \textbf{Constraints} \\ \hline
    All-Interval & \textit{alldifferent}, $x = |y-z|$\\
    Costas-Array & \textit{alldifferent}, $x - y = z$, $x - y \neq z - u$\\
    Haystacks    & $z + u \geq z$, $0 > (x-y) (z-u)$, $x \neq y$, $x=y$\\ \hline
  \end{tabular}}
  \vspace{3pt}
  \caption{Intensional Constraints Used}
  \label{table:constraints}
  \vspace{-18pt}
\end{table}

\begin{figure*}[tb]
\center{
    \begin{minipage}[b]{150pt}
    \includegraphics[width=\textwidth, scale=0.75]{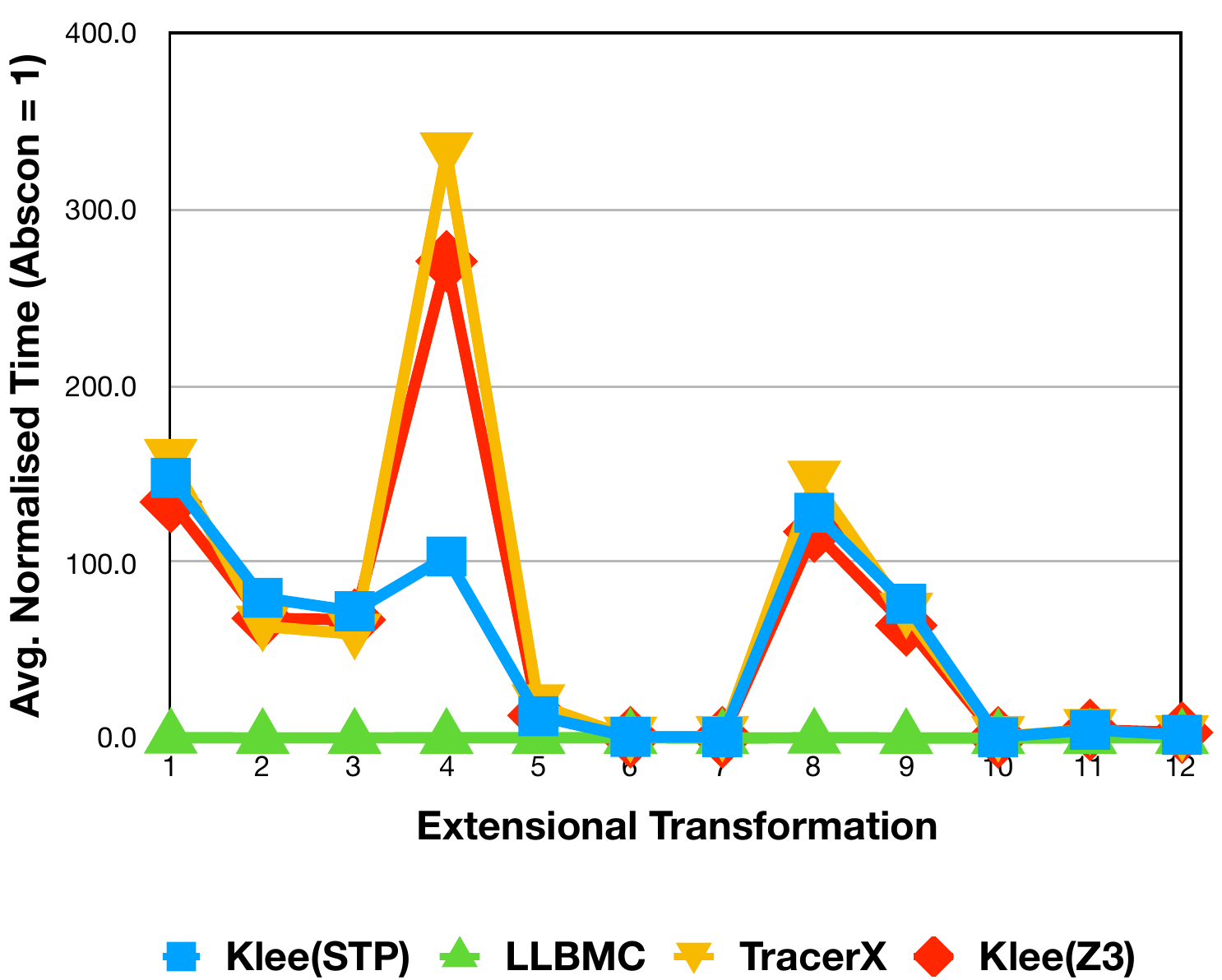}
    \vspace*{-20pt}
    \center{\small(a) AIM-100 Satisfiable}
    \includegraphics[width=\textwidth, scale=0.75]{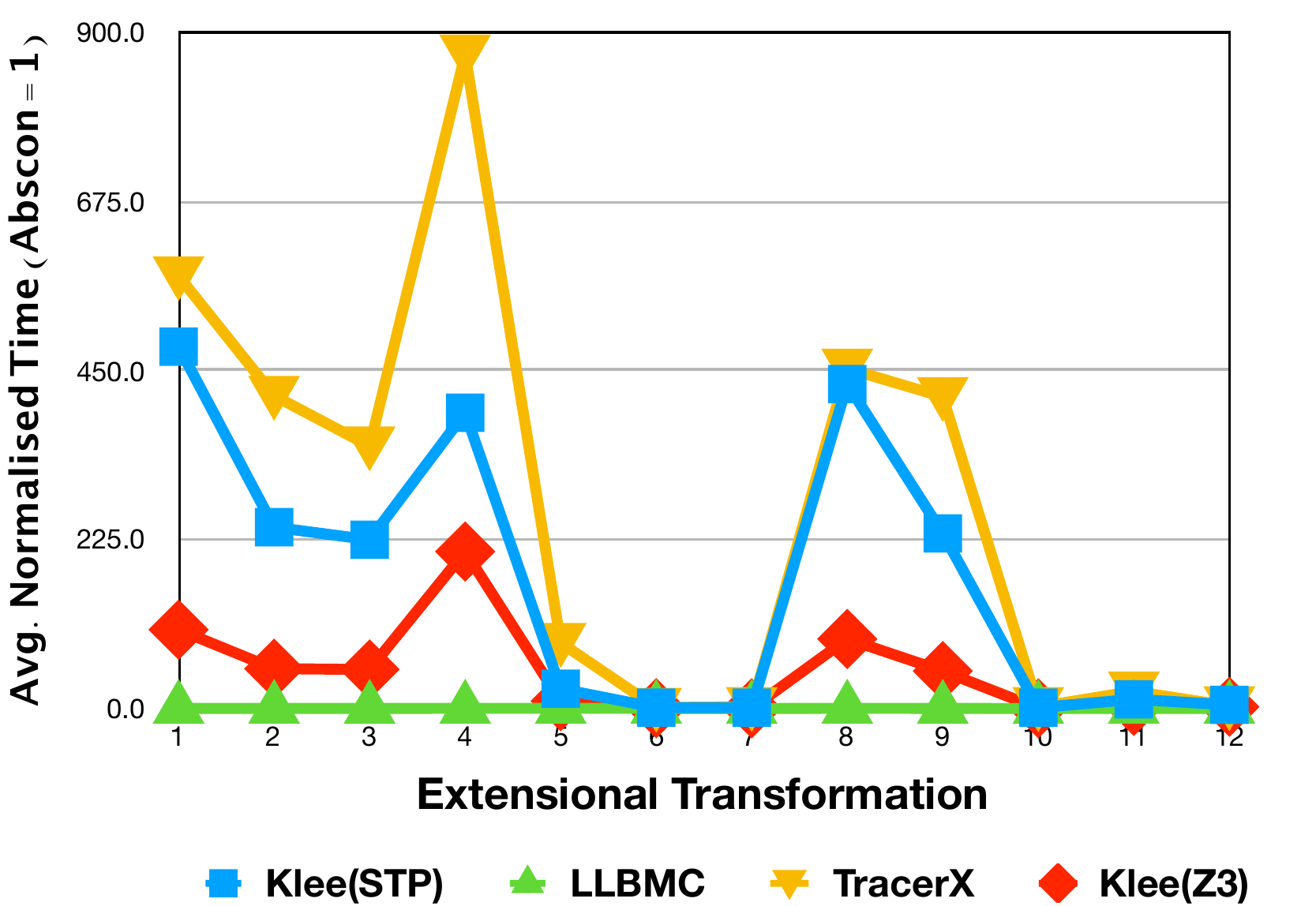}
    \vspace*{-20pt}
    \center{\small(b) AIM-200 Satisfiable}
  \end{minipage}
  \begin{minipage}[b]{150pt}
    \includegraphics[width=\textwidth, scale=0.75]{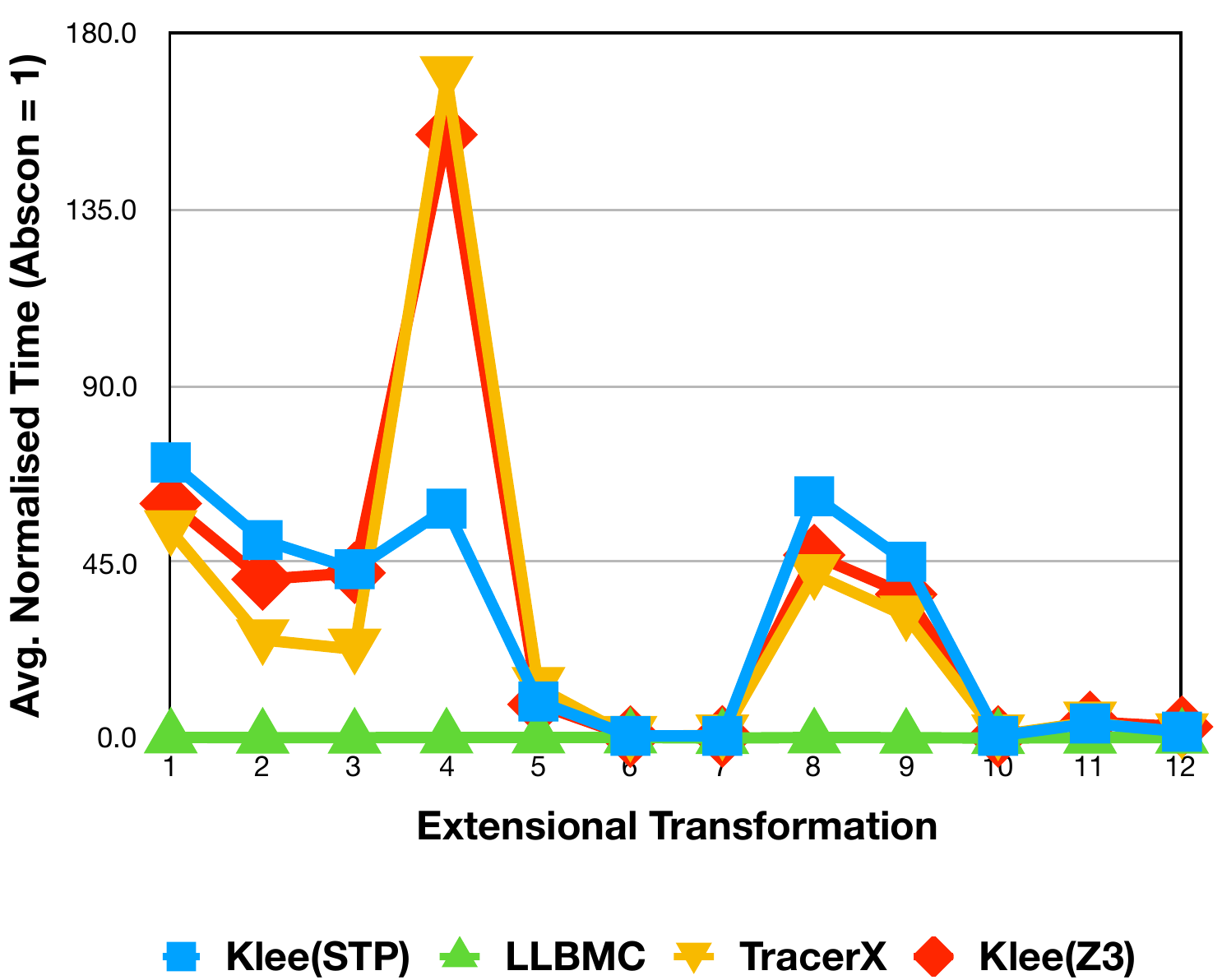}
    \vspace*{-20pt}
    \center{\small(c) AIM-100 Unsatisfiable}
    \includegraphics[width=0.98\textwidth, scale=0.75]{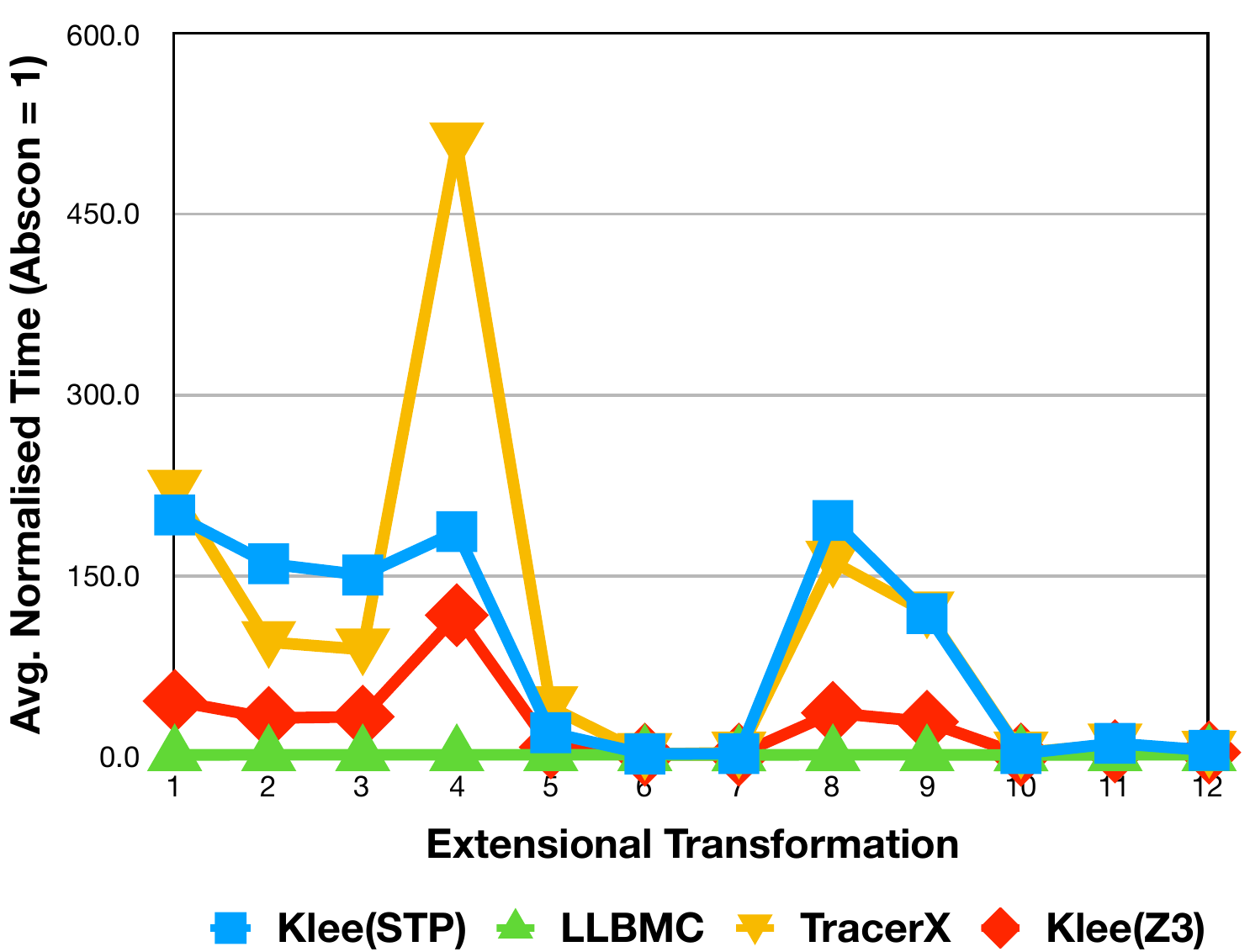}
    \vspace*{-20pt}
    \center{\small(d) AIM-200 Unsatisfiable}
    \vspace{-5pt}
  \end{minipage}
}
  \caption{Extensional Problems Transformation Robustness}
  \label{fig:extensional-robustness}
  \vspace{-12pt}
\end{figure*}

\begin{figure}[tb]
\begin{center}
  \includegraphics[width=0.4\textwidth, scale=0.65]{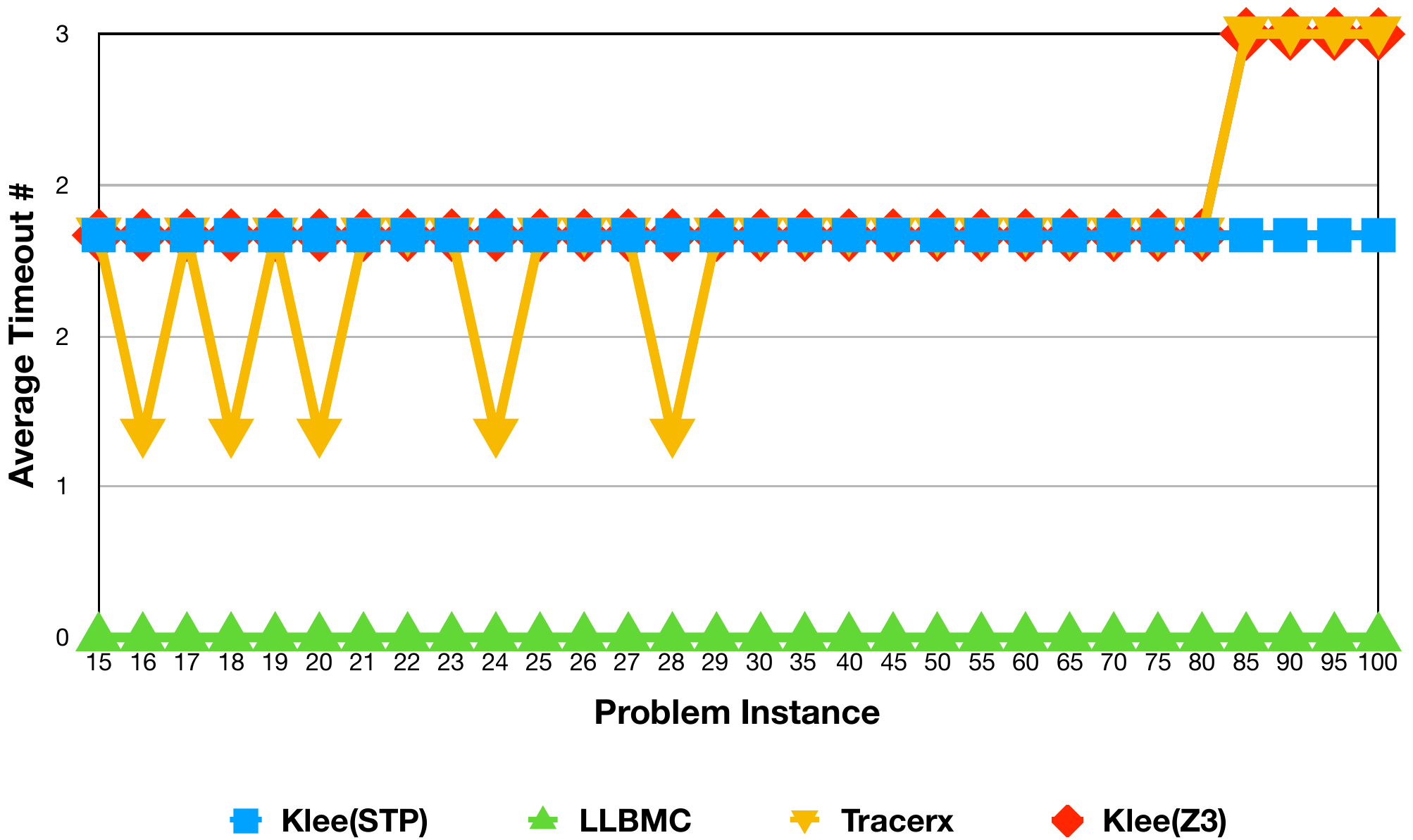}
\end{center}
\vspace{-12pt}
\caption{Extensional Problem Scalability - Dubois}
\label{fig:extensional-scalability}
 \vspace{-15pt}
\end{figure}

\begin{figure}[tb]
  \center{\includegraphics[width=0.28\textwidth, scale=0.35]{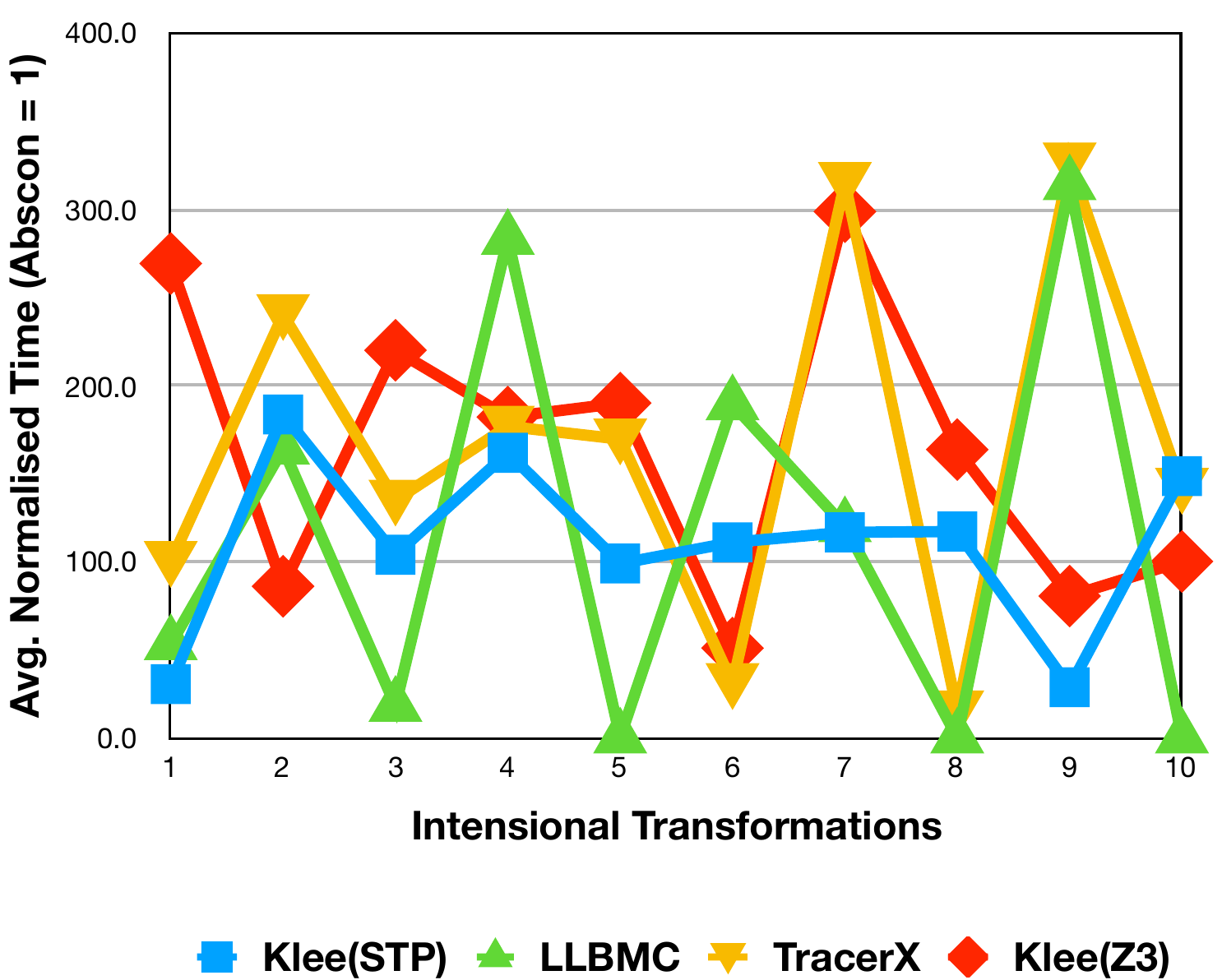}}

  \vspace*{-3mm}
  \center{\small(a) All-Interval}
  \vspace*{-3mm}

  \center{\includegraphics[width=0.28\textwidth, scale=0.35]{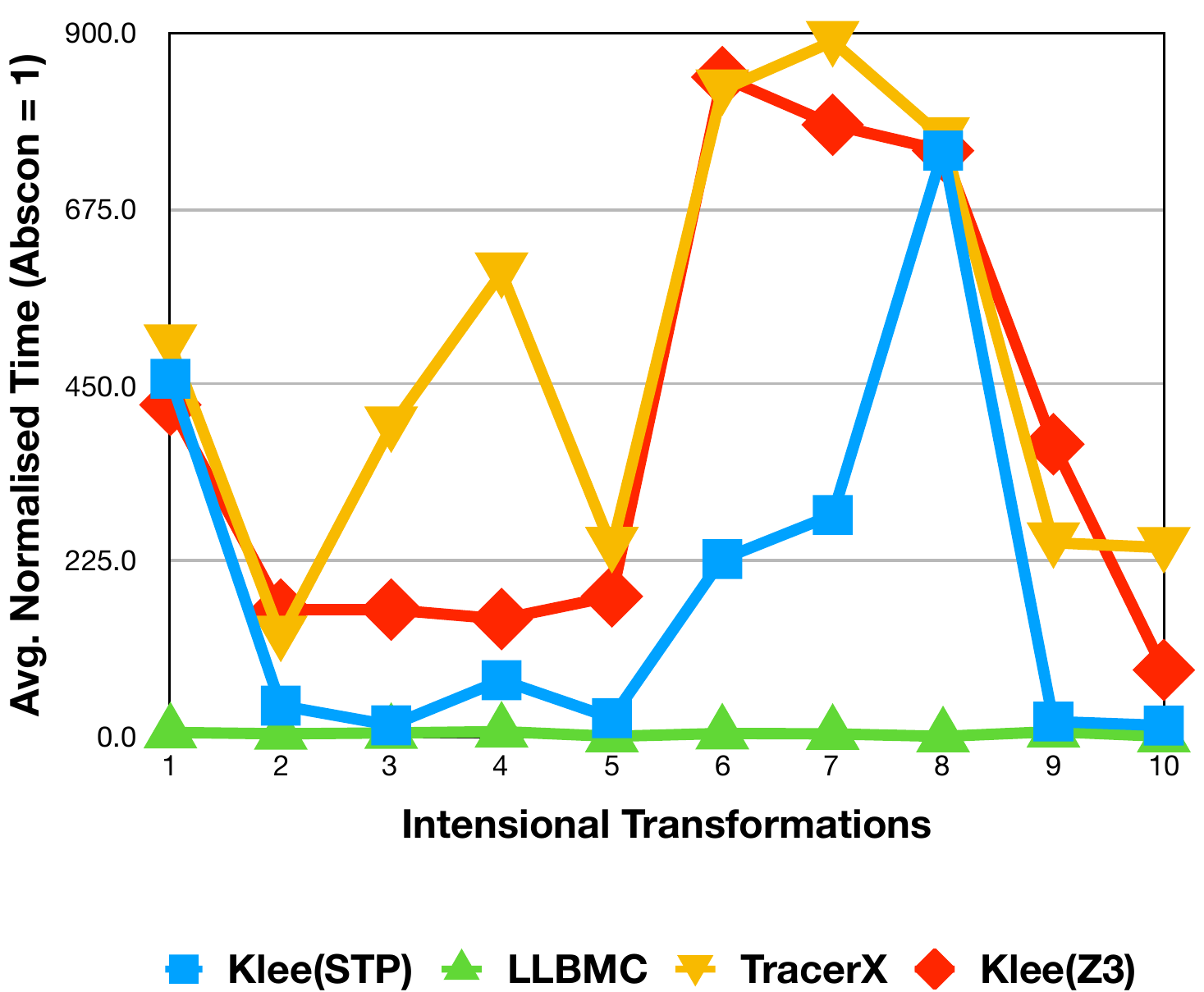}}

  \vspace*{-3mm}
  \center{\small(b) Costas-Array}
  \vspace*{-3mm}

  \center{\includegraphics[width=0.28\textwidth, scale=0.35]{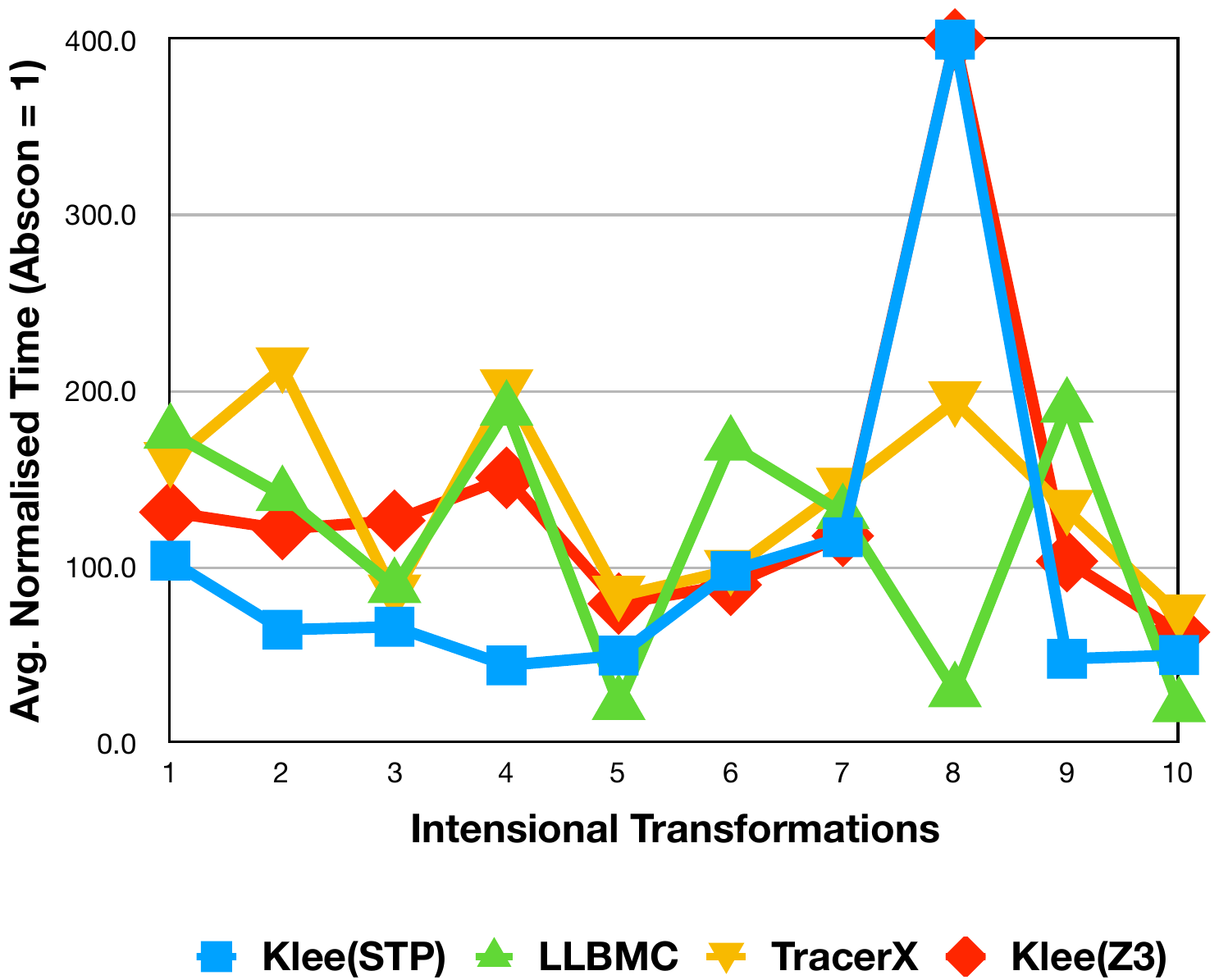}}

  \vspace*{-3mm}
  \center{\small(c) Haystacks}
  \vspace*{-2mm}

  \caption{Intensional Problems Transformation Robustness}
  \label{fig:intensional-robustness}
 \vspace{-15pt}
\end{figure}

\begin{figure}[tb]
  \center{\includegraphics[width=0.28\textwidth, scale=0.35]{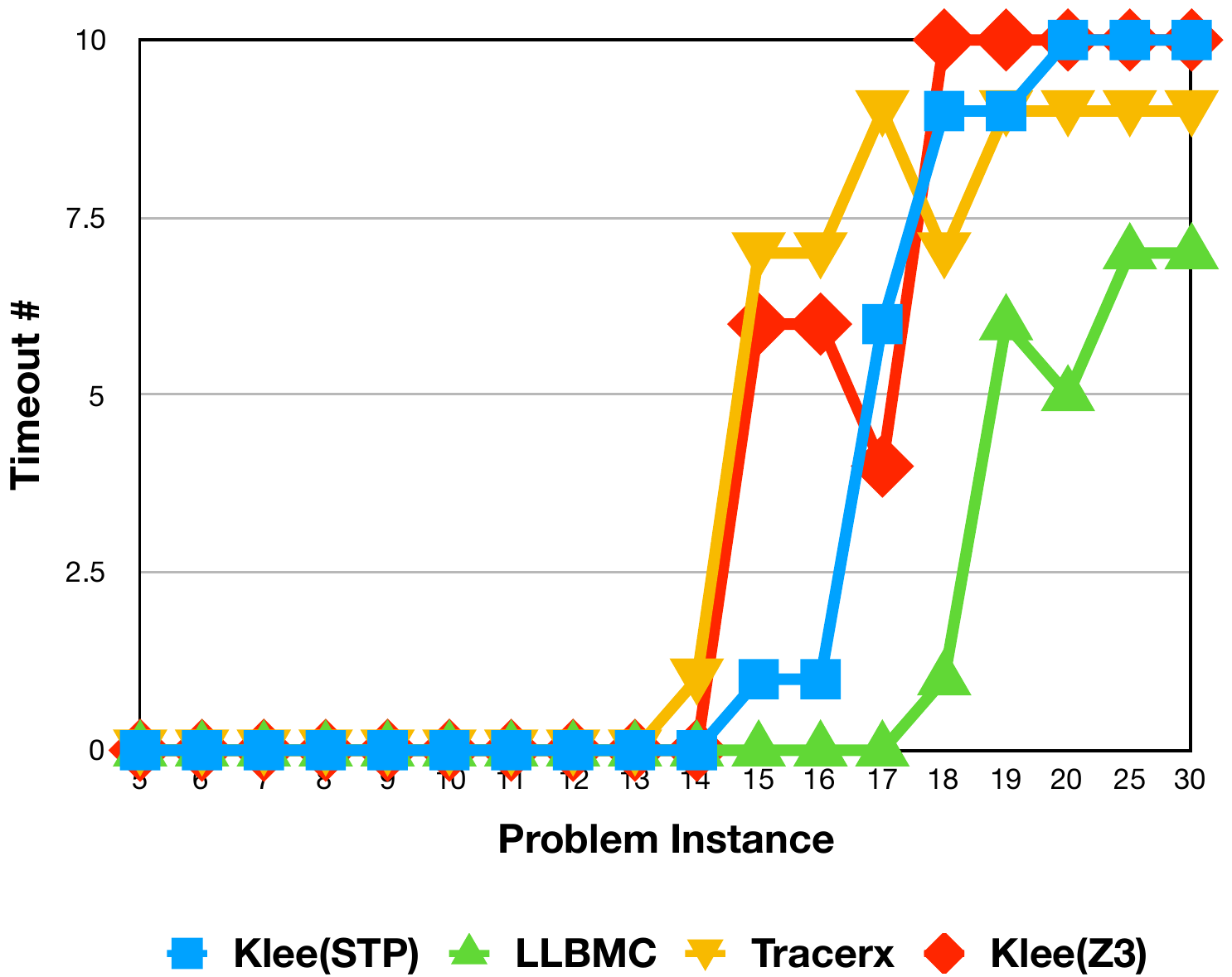}}

  \vspace*{-2mm}
  \center{\small(a) All-Interval}
  \vspace*{-2mm}

  \center{\includegraphics[width=0.28\textwidth, scale=0.35]{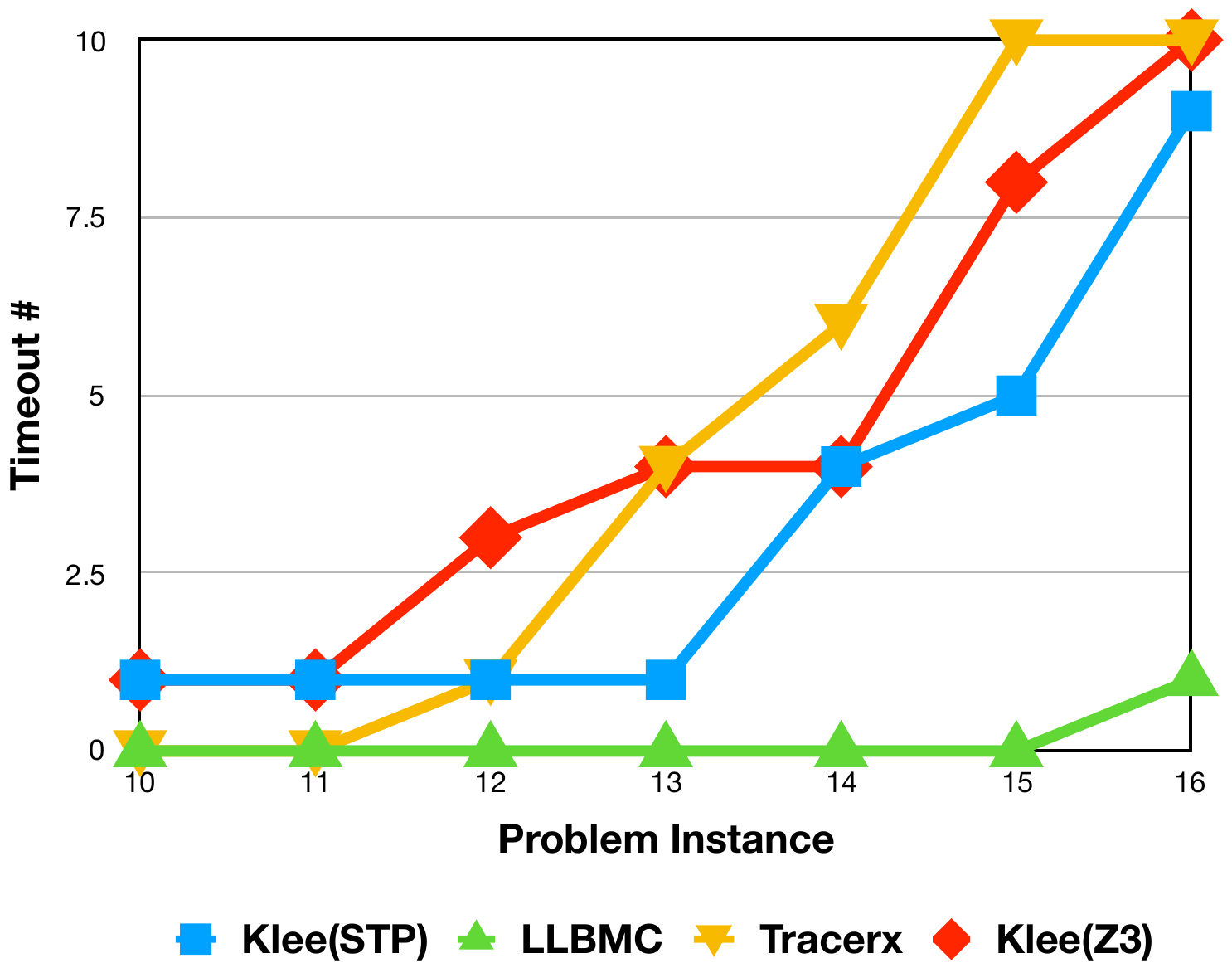}}

  \vspace*{-2mm}
  \center{\small(b) Costas-Array}
  \vspace*{-2mm}

  \center{\includegraphics[width=0.28\textwidth, scale=0.35]{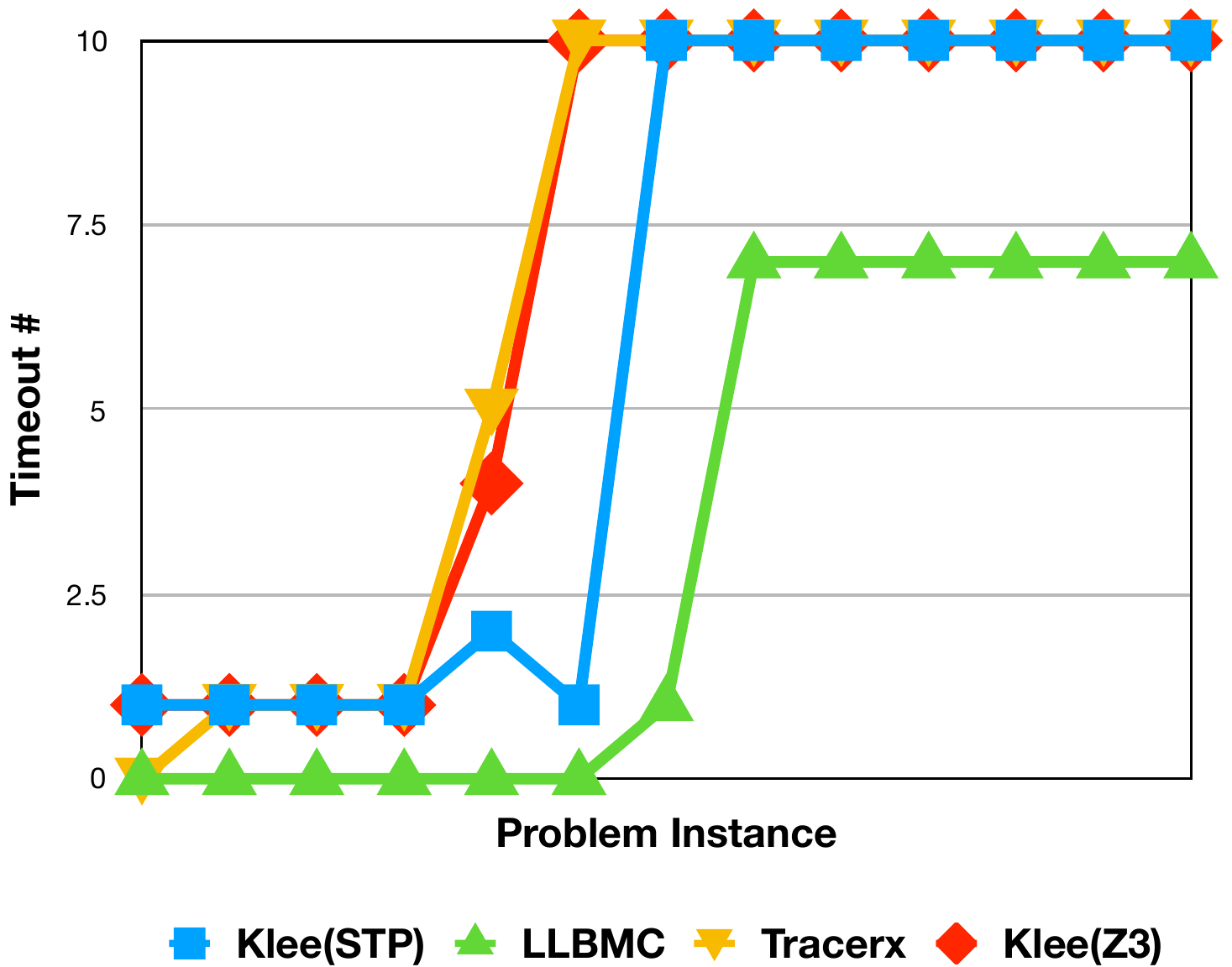}}

  \vspace*{-2mm}
  \center{\small(c) Haystacks}
  \vspace*{-2mm}

  \caption{Intensional Problems Scalability}
  \label{fig:intensional-scalability}
\vspace*{-5mm}
\end{figure}

\vspace{-1mm}
\subsection{Evaluation}

Figure \ref{fig:extensional-robustness} investigates transformation
robustness for the extensional problems.  We use average timings
for analyzing the C code normalized to AbsCon solving time of the XCSP instance.
% and in the case where most instances timeout,
% {\it timeout counts}, as in Figure \ref{fig:extensional-robustness} (b).
Different transformations result in a different number of paths explored
for the KLEE-based tools. We see that the KLEE-based
tools are sensitive to the transformations and exhibit considerable variation, up to 2
orders of magnitude slower for certain transformations. They also
exhibit a similar pattern in the transformation running times between
AIM-100 and AIM-200 instances.
%\RY{dubois not shown but similar}
Due to lack of space, 
results for Dubois are not shown, but it has a similar
pattern to AIM instances.
The timings show that there can be a complex relationship between the solver and
performance of transformation. 
Sometimes KLEE/Z3 and Tracer-X/Z3 are 
slower than KLEE/STP but in others, it is faster (AIM-200 Version 4),
showing the importance of the solver effect.
(We remark that KLEE and STP are co-designed 
while Z3 is independent).
Greater differences can be seen in the intensional benchmarks.
We can see that generally, the greater the grouping (Table \ref{table:features}
Grouped column),
the better the performance, in order of {\it no} to {\it yes} to {\it all}. 
There is a large difference between 
logical versus bitwise operators, e.g., Version 1 versus Version 4.

We see that LLBMC is significantly more robust under all
transformations. This may be because LLBMC 
creates a single (SMT) constraint problem.
The LLBMC runtime relative to Abscon in
Figure \ref{fig:extensional-robustness} for AIM-100 is between
0.15-0.22 and for AIM-200 between 0.17-0.26, i.e., LLBMC is faster.

\begin{comment}
We observe that while the behavior of SATisfiable problems
(Figure \ref{fig:extensional-robustness} (a) and (b)) is
similar to UNSATisfiable problems 
(Figure \ref{fig:extensional-robustness} (c) and (d)),
there are some differences. The difference between Tracer-X and KLEE
with the Z3 solver are very close with UNSAT problems but more
differentiated with SAT problems.
There is also more variation in robustness with the UNSAT problems; for example, the Extensional Version 4 problems are more difficult
for KLEE than LLBMC.
\end{comment}

%we make the constraints load the variable value from a dynamically-allocated memory before being evaluated.  which seems to trigger an expensive solving procedure within the STP solver used by LLBMC. 

The results for the transformation robustness of intensional problems is
shown in Figure \ref{fig:intensional-robustness}. 
\begin{comment}
A similar tendency
where KLEE/STP is faster than the Z3-based systems can be seen.  LLBMC
is also more robust than KLEE but less robust than with the
extensional benchmarks.  
\end{comment}
While LLBMC is still often better than KLEE, there is a larger variance.
For example, All-Interval Version 9, KLEE(STP) is better than LLBMC 
but Tracer-X, which is KLEE-based, is similar to LLBMC.
In the extensional code, the effort in Tracer-X to compute interpolants turned out
to be wasted, but the value in the reduction of paths can be seen here where Tracer-X can
outperform KLEE(STP/Z3). Here, STP is generally faster than Z3 compared to the extensional
problems which only use equals or not-equals constraints.
Haystacks shows an interesting solver issue, the only transformation difference between Version 3 and 8 is
the {\bf if} versus {\bf assume}---one might think that the {\bf assume} 
(more direct solver usage)
should handle the constraints better, but it turns out to be worse. 
We conjecture this may be due to differences in the handling
of the short-circuiting from the logical operators.
Tracer-X behaves better for these instances.

Figures \ref{fig:extensional-robustness} and
\ref{fig:intensional-robustness} use running time normalized to
AbsCon running time. 
This shows that the KLEE-based symbolic execution tools are significantly slower than running the original CSP problem with the AbsCon solver. 
Note that this is not intended to be a fair solver comparison, all the
systems use some solvers, but the solvers are used in different ways
and the solvers themselves employ different techniques.
LLBMC is often faster than AbsCon for the AIM problems. 
One possible reason is that the AIM instances are SAT problems and the
AIM instances are known to be easy for modern SAT solvers. 
LLBMC uses STP, which
in turn, uses the MiniSat SAT solver \cite{minisat} can be expected to give good results.
The AbsCon solver uses generalized arc consistency
for the extensional constraints and does not employ learning.
The performance of SAT solvers, on the other hand,
is highly dependent on the use of clause learning.
The intensional benchmarks show that STP solver performance is different
from the extensional benchmarks, now STP is generally faster than Z3
but All-Interval Version 2 shows this is not always the case.

% \RY{dubois also goes here}
We investigate tool scalability with extensional (Dubois) and intensional problems where
there is a size parameter.
The results are shown in Figures \ref{fig:extensional-scalability} and \ref{fig:intensional-scalability}.  
In these graphs, 
a greater problem number (x-axis)
indicates a larger size parameter and the y-axis gives the number
of timeouts (more timeout is worse). As can be observed, LLBMC is more
scalable than the others, with KLEE/STP slightly more scalable than
the two KLEE-based tools with the Z3 solver.
We can see for the extensional problems that although Tracer-X was generally slower
in the AIM benchmarks, it can solve more problems compared with KLEE(STP/Z3).

\begin{comment}
We see that the variance due to solvers is not a simple relationship,
see Figure \ref{fig:intensional-scalability}, where STP is generally
faster but is slower for Intensional Version 8 for Haystacks
instances.
STP being sometimes slower than Z3 also happens for some other problems.
\end{comment}

In summary, we see that LLBMC generally executes the
benchmarks faster than other tools. It is known in the literature that
the invocation of the constraint solver is expensive 
(e.g., \cite{zhang12speculative}). During path enumeration, KLEE-based tools
invoke the constraint solver multiple times to determine the
feasibility of control-flow branches. Compared to the KLEE-based
tools, LLBMC invokes the solver only once with a single global (much larger)
problem.  The versions run on KLEE-based systems which
correspond most closely to LLBMC solver usage are: Extensional Versions
6 and 12 and Intensional Versions 5 and 10.  As can be seen from Figures
\ref{fig:extensional-robustness} and \ref{fig:intensional-robustness},
these versions show much lower execution times for the KLEE-based tools compared with the other transformations.

Our experiments may give a different impression from conventional use
of KLEE on real-world C programs, e.g., KLEE may be expected to work on
small programs but may not scale to large ones; however, the Dubois C programs are 
small (hundreds of lines) and still timeout.
This may be because solving and path
explosion is more pronounced in our ``benchmark code'' versus
``typical code''.
% The results of our preliminary study 
% may be quite different when benchmarking KLEE tools
% on easier constraint cases which arise in many real-world C programs.
However, the choice of artificial code is deliberate---to allow us to study
the tradeoffs and interactions between the solver and path exploration.
Real-world programs may pose less of a challenge to this aspect,
but can still pose challenges in other aspects, such as system calls.
The results show that the execution times of the KLEE-based
tools on the benchmarks depend heavily on the solver used (STP or
Z3) and are significantly affected by the C constructs (from the transformations).
Although both LLBMC and KLEE/STP use the same solver, the timings 
differ substantially, which suggests that KLEE-based symbolic execution
has considerable room for optimizations.
Our benchmarks are designed so that the reasoning on the program execution
paths are nontrivial.
Tracer-X uses interpolants which can prune (irrelevant) execution paths,
however, in our benchmarks, interpolants can have some benefit but the cost
of interpolants is also a drawback.
This suggests that the use of interpolants may not help so much in the harder path exploration cases arising in our benchmarks. 

% Our data is publicly available~\cite{verma18cspc}.
% As the execution paths in our benchmarks are less easy to optimize, it appears that more sophisticated search techniques such as in Tracer-X which themselves have a cost are not sufficient, and not from the sophisticated search technique used, since both KLEE/Z3 and Tracer-X have similar running times.

%% \begin{figure*}[t]
%%   \begin{minipage}{235pt}
%%   \includegraphics[width=230pt]{aim100-no-timeout-sat-time.png}
%%   \center{(a) AIM-100 Satisfiable}
%%   \includegraphics[width=230pt]{aim200-timeout-count-sat-time.png}
%%   \center{(b) AIM-200 Satisfiable}
%%   \end{minipage}
%%   \begin{minipage}{235pt}
%%   \includegraphics[width=230pt]{aim100-no-timeout-unsat-time.png}
%%   \center{(c) AIM-100 Unsatisfiable}
%%   \includegraphics[width=230pt]{aim200-no-timeout-unsat-time.png}
%%   \center{(d) AIM-200 Unsatisfiable}
%%   \end{minipage}
%%   \center{\includegraphics[width=200pt]{legend.png}}
%%   \caption{Extensional Problems Execution Time}
%%   \label{fig:extensional-time}
%% \end{figure*}

\subsection{Discussion}

We summarize how the experimental results 
can help address the performance of
symbolic execution with the following research questions:
\begin{itemize}
\item Q1:{\it What kinds of programs are easier/harder to analyze with symbolic execution?}
The experiments show that KLEE/STP, KLEE/Z3, and Tracer-X/Z3 are not
robust under different transformations, thus, programs coded in
a particular way will be harder to analyze.

\item Q2:{\it Do different constraints change the difficulty of the symbolic execution?}
The benchmarks use various constraints.  The extensional benchmarks use simple equality constraints of the form $x = c$ or $x \neq c$, while the intensional benchmarks have more complex constraints, e.g. $0 > (x-y) (z-u)$.
There is a larger variation in the overall robustness with the 
intensional benchmarks.

\item Q3:{\it Are there differences in overall solver performance between different solvers?}
The experiments show solver choice is significant, e.g., between STP and Z3; however, this relationship is complex and may not be simple to
predict. From anecdotal experience, KLEE/STP is expected to outperform KLEE/Z3; however,
for certain kinds of code, KLEE/Z3 can be much faster.
\end{itemize}

Overall, the results suggest that considerable improvements to symbolic
execution tools may be possible, such as better solver integration and improved solver reasoning.
While Tracer-X was in many cases slower than KLEE, possibly due to the
interpolation overhead, there were cases where
it was faster, which suggests that smarter removal of execution paths
is beneficial but much research remains to be done 
as shown by the overall results. 
As our transformations can be considered a kind of pre-processing
step, a smart pre-processing step may be another direction for improving
symbolic execution. This may be analogous to some constraint solvers which 
perform extensive pre-processing, for example, the pre-processing phase
in the CPLEX MIP solver is sufficient to solve some problems without going
to the actual solver \cite{achterberg}.
As far as we are aware, such more advanced constraint processing ideas 
have not been employed in symbolic execution.

We highlight that the objective of this paper is to build benchmarks
suitable to improve the core reasoning of symbolic execution.
So while our benchmarks generally have bounded model checking (LLBMC)
being faster, this is partly an artifact because we have simple systematic
code which is loop-free. The comparison with LLBMC is also 
because many of the core reasoning components are similar to 
KLEE and Tracer-X, in particular,
the same underlying solver.
We found instances where KLEE/STP outperforms LLBMC.
For general and more complex code, symbolic execution has advantages over bounded model checking. The main limitation of bounded model checking
being the bound which also affects scalability if the bound is large.
Symbolic execution also has advantages when it comes to system calls and mixing
with concrete execution.
However, for our benchmarks, we want to deliberately focus on the solver,
path exploration and search, so these other factors are orthogonal to
our work.

While we have deliberately chosen simple transformations, more complex ones
are possible.
Rather than loop-free code,
extensional constraints can be transformed into a generic
table driven code with loops.
This may not work so well on LLBMC if the specified bound is exceeded
but would make little difference to KLEE.
We have chosen a simple code structure to focus on the solver,
path exploration and constraint handling mechanisms intrinsic to 
symbolic execution.

% \section{Discussion}
% \label{sec:discussion}

% \input{related}

\section{Conclusion}
\label{sec:conclusion}

We proposed finite-domain constraint satisfaction problems
(CSPs) for evaluating the effectiveness of the reasoning techniques in
symbolic execution tools.  We show how to encode CSP instances by transforming the problem into C programs using different C constructs.
Although multiple C programs are created, they are essentially equivalent from
the satisfiability and solution of the CSP perspective.
Preliminary testing with the KLEE, Tracer-X, and LLBMC tools with STP
and Z3 solvers show that the transformation used significantly affects
the performance of the tools even though the underlying reasoning
problem is the same.  
% The running times vary significantly per transformation which employs different C constructs and encoding of the CSP. 
The degree of sensitivity of symbolic execution tools to ``writing style'' of
the program has not been studied, and we feel is an important direction
for improved and more robust symbolic execution technology.

The use of logical operators common in C
has a significant effect compared to the use of bitwise operators. 
It suggests that a viable strategy may be to flip
the semantics of the code between bitwise and logical operators in cases where there are no side effects. 
This could give a performance boost---we are not aware
that this strategy has been tried before.

Ideally, as the programs are ``equivalent to the CSP instance'', we may expect performance from the tools to be not dissimilar since
the underlying reasoning techniques are also similar.
However, our benchmarks also show complex
and substantial differences.
% in solver running times simply by replacing the solver used for the same tool.  
There is also a large performance gap
between LLBMC and KLEE/STP even when both use the same underlying STP
solver, showing another direction for system improvement.  Although
our benchmarks are synthetic, precisely because they focus on the
basic core reasoning engine, we believe that benchmarks of these form
can help drive basic improvements to symbolic execution tools.
What we would like is that the analysis tools are more robust and more scalable.
We believe that if symbolic execution-based tools can perform better
on the kinds of benchmarks we proposed, this will carry forward to
improved overall analysis performance. In particular, for challenging code
such as cryptographic code, obfuscated code, and malware where program
analysis tools may not work well.

The results using Tracer-X suggest that optimizations which prune paths
can increase scalability but it also suggests that much more 
improvements are needed.
Future work may be to investigate redundant constraints as yet another dimension.
% In future work, one additional dimension to the benchmarking of symbolic execution will be to use CSP problems which have some amount of redundant constraints. This will give another dimension to evaluating how good is the solver and path exploration components.
Finally, we hope this can spur more synergy between techniques in constraint programming,
testing and program analysis.

\clearpage

\section*{Acknowledgement}
We acknowledge the support of R-252-000-592-112 and R-252-000-A39-112.
We would like to thank Andrew Santosa who helped with a preliminary 
version of this work. 

%% no space for ACK at the moment

\balance
\bibliographystyle{IEEEtran}
\bibliography{ShortStrings,xcsp3} 

\end{document}